\documentclass[final,3p,twocolumn,sort&compress]{elsarticle}

\usepackage{amsmath}
\usepackage{amssymb}
\usepackage[figurename=Fig.,labelsep=period]{caption}
\usepackage[colorlinks,citecolor=blue,urlcolor=blue,linkcolor=blue]{hyperref}
\usepackage{slashed}

\journal{Physics Letters B}

\begin{document}

\begin{frontmatter}



\title{A nontopological soliton in an $\mathcal{N} = 1$ supersymmetric gauge
       Abelian model}

\author[tusur]{A.Yu.~Loginov}
\ead{a.yu.loginov@tusur.ru}

\address[tusur]{Laboratory of Applied Mathematics and Theoretical Physics, Tomsk State
                University of Control Systems and Radioelectronics, 634050 Tomsk, Russia}
%\address{This article is registered under preprint number: arXiv:2108.05121 [hep-th]}

\begin{abstract}
A version of  $\mathcal{N}  =  1$   supersymmetric   scalar  electrodynamics is
considered here, and it is shown that  an  electrically  charged nontopological
soliton  exists in this model.
In addition  to  the  long-range  electric  field, the soliton also possesses a
long-range  scalar  field,  which leads to a modification  of  the intersoliton
interaction potential at large distances.
The supersymmetry of the model  makes  it  possible  to  express fermionic zero
modes of the soliton in terms of bosonic fields.
The properties of the nontopological soliton  are investigated using analytical
and numerical methods.
\end{abstract}

\begin{keyword}
nontopological soliton \sep electric charge \sep supersymmetry \sep fermionic zero modes
%\sep arXiv: 2108.05121

%\PACS 71.35.-y \sep 71.35.Lk \sep 71.36.+c


\end{keyword}

\end{frontmatter}

\section{Introduction}
\label{seq:I}

Many models of  field  theory have solutions  that describe spatially localised
and nonspreading  field  configurations   with  a  finite  energy \cite{Manton,
 Rubakov}.
Nontopological solitons \cite{lee_pang_1992}  represent   one   of  these field
configurations.
A necessary condition for the  existence  of  a nontopological  soliton  is the
symmetry of the corresponding field  model, which may be both global and local.
In addition, the  interaction  potentials  of  the  model  must  meet a certain
condition \cite{coleman_npb_1985, paccetti_npb_2001}.
The symmetry of the model  results  in  the  existence  of  a conserved Noether
charge.
The field configuration  of  a nontopological  soliton  is an extremum (minimum
or saddle point) of the  energy  functional  at  a  fixed  value of the Noether
charge, and this basic property  largely  determines  the other properties of a
nontopological soliton; in particular,  it  leads  to  the  characteristic time
dependence $\exp \left( - i\omega t \right)$ of a soliton field.

Nontopological solitons may  be  formed  during  a primordial phase transition,
thus making a contribution to various scenarios of the  evolution  of the early
Universe \cite{kus_plb_1997}.
Furthermore, they  may  play  an   essential   role   in  baryogenesis  via the
Affleck-Dine  mechanism  \cite{aff_dine_npb_1985},  and   are  considered to be
places where dark matter may be concentrated \cite{kus_shp_plb_1998}.

Some field  models  with  local   Abelian   symmetry   admit  the  existence of
electrically charged nontopological solitons.
First  described  in   Refs.~\cite{ros_1968_2, klee},   they  have  since  been
investigated in  many  other works (see, e.g., Refs.~\cite{lee_yoon_1991, anag,
 levi,  rduvlk_2008,   ardoz_2009,   tamaki_2014,   gulamov_2014,  br_prd_2014,
  hong_prd_2015, gulamov_2015, lshnir_2019, loginov_2020, lshnir_2022}).
The properties  of  electrically  charged  solitons  differ  significantly from
those of solitons without  an  electric  charge;  in particular,  the  electric
charge and the energy  of a nontopological soliton cannot  be arbitrarily large
in the  general  case \cite{klee, gulamov_2015}.
In addition, an electrically  charged nontopological  soliton can exist only if
the gauge coupling constant does  not exceed  some  maximum  value \cite{klee}.

The main goal of this work  is  to study a non-topological soliton in a version
of $\mathcal{N}  =  1$  supersymmetric  scalar electrodynamics.
The interaction  potential  of   this   model   is   expressed  in  terms  of a
superpotential, which  leads  to  relations  between  the nonlinear interaction
constants.
In addition,   the   superpotential   largely   determines   the   form  of the
scalar-fermion interaction.
The  requirements  of  renormalisability  and  gauge  invariance  impose severe
restrictions on  the  form  of  the  superpotential, all of which significantly
reduces the number of  model parameters compared to the nonsupersymmetric case.

Throughout this paper, we use the natural units $c = 1$, $\hbar = 1$.
The metric  tensor   and   the   Dirac  matrices   are   defined   according to
Ref.~\cite{Weinberg_III}.

\section{Lagrangian and field equations of the model} \label{seq:II}

The $\mathcal{N} = 1$ supersymmetric gauge  model  under consideration includes
three left-chiral matter  superfields $\Phi_{-1}$, $\Phi_{0}$, and $\Phi_{+1}$,
and one Abelian gauge superfield $V$.
The left-chiral  superfield  $\Phi_{n}$  contains  two  components: the complex
scalar field $\phi_{n}$ and the left-hand   Dirac  spinor  field  $\psi_{n L}$.
Written in the Wess-Zumino gauge,  the  gauge  superfield $V$ also contains two
components: the Abelian gauge field  $A_{\mu}$  and  the  Majorana spinor field
$\lambda$.
The superfields $\Phi_{n}$ and $V$ also contain auxiliary fields, but these can
be expressed in terms of the above mentioned physical fields.

The Lagrangian of the model takes the form
\begin{flalign}
& \mathcal{L} = -\frac{1}{4}F_{\mu\nu}F^{\mu\nu}-\sum\limits_{n}\left(
D_{\mu}\phi_{n}\right)^{\ast }D^{\mu}\phi_{n}-V\left(\phi\right)
\nonumber  \\
&-\frac{1}{2}\bar{\lambda}\gamma^{\mu}\partial_{\mu}\lambda
-\sum\limits_{n}\overline{\psi_{nL}}\gamma^{\mu}D_{\mu}\psi_{nL}
\nonumber  \\
&-\frac{1}{2}\sum\limits_{nm}\left\{f_{nm}\left(\psi_{nL}^{\text{T}
}\epsilon\psi_{mL}\right)+f_{nm}^{\ast}\left(\psi_{nL}^{\text{T}
}\epsilon \psi_{mL}\right)^{\ast }\right\}
\nonumber  \\
&+i\sqrt{2}\sum\limits_{n}q_{n}\left\{\phi_{n}\left(\overline{\psi_{nL}}
\lambda\right)-\phi_{n}^{\ast}\left(\bar{\lambda}\psi_{nL}\right)
\right\}.                                                          \label{II:1}
\end{flalign}
In Eq.~(\ref{II:1}), the matrix $\epsilon = -i \gamma_{0}\gamma_{2}\gamma_{5}$,
the Latin indices $n$ and $m$ run over the set  $[-1, 0, 1]$, and the covariant
derivatives
\begin{subequations}                                               \label{II:2}
\begin{flalign}
D_{\mu}\phi_{n} &=\partial_{\mu}\phi_{n}-iq_{n}A_{\mu}\phi_{n},   \label{II:2a}
\\
D_{\mu}\psi_{n} &=\partial_{\mu}\psi_{n}-iq_{n}A_{\mu}\psi_{n},   \label{II:2b}
\end{flalign}
\end{subequations}
where $q_{n} = n e$ are  the  Abelian  charges  of  the  left-chiral superfield
$\Phi_{n}$.
To avoid  $U(1)$-$U(1)$-$U(1)$  and $U(1)$-graviton-graviton anomalies, the sum
of the $U(1)$  quantum  numbers  of  all left-chiral superfields and the sum of
their cubes should vanish, which is obviously true in our case.
The  field-dependent  coefficients  $f_{nm}$  and   the  interaction  potential
$V\left(\phi\right)$ are expressed in terms of the superpotential
\begin{equation}
f\left(\phi\right)=m\phi_{-1}\phi_{+1}+g\phi_{-1}\phi_{0}\phi_{+1},\label{II:3}
\end{equation}
where $m$ is a mass parameter and $g$ is a coupling constant.
The coefficients  $f_{nm} = \partial^{2}f/ \partial\phi_{n} \partial \phi_{m}$,
and the interaction potential
\begin{flalign}
V\left( \phi \right)  &=\sum\limits_{n}\left\vert\partial f/\partial\phi_{n}
\right\vert^{2}+\frac{1}{2}\Bigl(
\sum\limits_{n}q_{n}\phi_{n}^{\ast}\phi_{n}\Bigr)^{2}               \nonumber
 \\
&=\left\vert m + g \phi_{0} \right\vert^{2}
\left(\left\vert \phi_{-1}\right\vert^{2} +
\left\vert \phi_{+1}\right\vert ^{2}\right)                         \nonumber
 \\
&+g^{2}\left\vert \phi_{-1}\right\vert^{2}\left\vert \phi
_{+1}\right\vert^{2}                                                \nonumber
 \\
&+\frac{e^{2}}{2}\left( \left\vert \phi_{+1}\right\vert
^{2}-\left\vert \phi _{-1}\right\vert ^{2}\right)^{2}.             \label{II:4}
\end{flalign}

The field equations of model (\ref{II:1}) have the form
\begin{equation}
\partial_{\mu }F^{\mu \nu }=j^{\nu },                             \label{II:5}
\end{equation}
\begin{flalign}
&D_{\mu }D^{\mu }\phi _{n}-\frac{\partial V}{\partial \phi _{n}^{\ast }}-
\frac{1}{2}\sum\limits_{k' m'}f_{k' m' n}^{\ast}
\left(\psi_{k'L}^{\text{T}}\epsilon \psi_{m'L}\right) ^{\ast}       \nonumber
 \\
&-i\sqrt{2}q_{n}\left( \overline{\lambda }\psi _{nL}\right) = 0,   \label{II:6}
\end{flalign}
\begin{equation}
\slashed{D}\psi_{nL}-\sum\limits_{m^{\prime }}f_{nm^{\prime
}}^{\ast}\epsilon
\left(\overline{\psi_{m^{\prime}L}}\right)^{\text{T}}
-i\sqrt{2}q_{n}\phi_{n}\lambda_{R} = 0,                            \label{II:7}
\end{equation}
\begin{equation}
\slashed{\partial}\lambda +i\sqrt{2}\sum\limits_{m^{\prime
}}q_{m^{\prime }}\left\{ \phi _{m^{\prime }}\epsilon \left( \overline{\psi
_{m^{\prime }L}}\right) ^{\text{T}}+\phi _{m^{\prime }}^{\ast }\psi
_{m^{\prime }L}\right\} = 0,                                       \label{II:8}
\end{equation}
where the  coefficients  $f_{k m n}  =  \partial^{3}f/\partial\phi_{k} \partial
\phi_{m} \partial \phi_{n}$ and the electromagnetic current
\begin{equation}
j^{\nu}=i\sum\limits_{n}q_{n}\phi_{n}^{\ast}\overset{\longleftrightarrow
}{D^{\nu}}\phi_{n}-i\sum\limits_{n}q_{n}\overline{\psi_{nL}}\gamma^{\nu
}\psi_{nL}.                                                        \label{II:9}
\end{equation}
Later on, we shall also  need  the  expression  for  the  energy  density of an
electrically charged bosonic field configuration of the model
\begin{flalign}
\mathcal{E} &=\frac{1}{2}E_{i}E_{i}+\sum\limits_{n}\left\{\left(
D_{t}\phi_{n}\right)^{\ast}D_{t}\phi_{n}\right.                     \nonumber
  \\
&\left. +\left(D_{i}\phi_{n}\right)^{\ast}D_{i}\phi_{n}\right\}
+ V\left(\phi\right),                                             \label{II:10}
\end{flalign}
where $E_{i} = F_{i 0}$  are  the  components  of  the electric field strength.

\section{Ansatz and some properties of the nontopological soliton}
\label{seq:III}

The model (\ref{II:1}) can be viewed as the Abelian gauge version of a model of
the Wess-Zumino type \cite{wz_npb_1974}.
In Ref.~\cite{lgn_pan_2010}, it  was shown that for superpotentials of the type
in Eq.~(\ref{II:3}),  these  models   admit   the  existence  of nontopological
solitons.
It follows from  continuity  considerations  that  nontopological  solitons can
also exist in gauge model (\ref{II:1}), at least  for sufficiently small values
of the gauge coupling constant $e$.

Let us define the shifted field $\varphi_{0}\left(\mathbf{x},t\right) = mg^{-1}
+\phi_{0}\left(\mathbf{x},t\right)$.
To find  a  nontopological  soliton  solution,  we  shall  use  the spherically
symmetrical ansatz:
\begin{subequations}                                              \label{III:1}
\begin{flalign}
\phi _{+1}\left( \mathbf{x},t\right) & = 2^{-\frac{1}{2}}
\exp \left(-i\omega t\right)f_{+1}\left( r\right),               \label{III:1a}
  \\
\phi _{-1}\left( \mathbf{x},t\right) & = 2^{-\frac{1}{2}}
\exp \left( i\omega t\right) f_{-1}\left( r\right),              \label{III:1b}
  \\
\varphi_{0}\left( \mathbf{x},t\right) & = 2^{-\frac{1}{2}}
(\chi_{1}\left(r\right)+i\chi_{2}\left( r\right)),               \label{III:1c}
  \\
A^{\mu }\left( \mathbf{x},t\right) & =\left( \Phi \left( r\right)
,\,0\right).                                                     \label{III:1d}
\end{flalign}
\end{subequations}
The energy density  (\ref{II:10}),  written   in  terms of the ansatz functions
(\ref{III:1}), takes the form
\begin{flalign}
\mathcal{E} &= \frac{1}{2}\Omega^{2}\left(f_{-1}^{2}+f_{+1}^{2}\right) +
\frac{1}{2} \Phi^{\prime 2}                                         \nonumber
 \\
& +\frac{1}{2}\left(f_{-1}^{\prime 2}+f_{+1}^{\prime 2}+\chi_{1}^{\prime 2}
+\chi_{2}^{\prime 2}\right) + V,                                  \label{III:2}
\end{flalign}
where the interaction potential
\begin{flalign}
V &= \frac{g^{2}}{4}\left( f_{-1}^{2}+f_{+1}^{2}\right)
\left(\chi_{1}^{2}+\chi _{2}^{2}\right)                             \nonumber
  \\
&+\frac{g^{2}}{4}f_{-1}^{2}f_{+1}^{2}+\frac{e^{2}}{8}
\left(f_{+1}^{2}-f_{-1}^{2}\right)^{2},                           \label{III:3}
\end{flalign}
the  function  $\Omega \left( r \right) = \omega - e\Phi \left( r \right)$, and
the prime indicates the derivative with respect to $r$.
The Lagrangian   density   $\mathcal{L}$   differs   from   the  energy density
$\mathcal{E}$ only  in  regard  to  the sign of the terms in the second line of
Eq.~(\ref{III:2}).
The electromagnetic  current  of  spherically  symmetrical  field configuration
(\ref{III:1}) is
\begin{equation}
j^{\nu}=\left(e \Omega \left(f_{-1}^{2}+f_{+1}^{2}\right), 0, 0, 0\right).
                                                                  \label{III:4}
\end{equation}

Substituting ansatz (\ref{III:1}) into  the  bosonic  parts  of field equations
(\ref{II:5})  and (\ref{II:6}),  we  obtain  a system of nonlinear differential
equations for the ansatz functions:
\begin{equation}
\Omega ^{\prime \prime }+\frac{2}{r}\Omega ^{\prime }-e^{2}\left(
f_{-1}^{2}+f_{+1}^{2}\right) \Omega =0,                           \label{III:5}
\end{equation}
\begin{equation}
f_{\pm 1}^{\prime \prime }+\frac{2}{r}f_{\pm 1}^{\prime }+\frac{\partial U}
{\partial f_{\pm 1}}=0,                                           \label{III:6}
\end{equation}
\begin{equation}
\chi_{1,2}^{\prime \prime}+\frac{2}{r}\chi_{1,2}^{\prime}+\frac{\partial U}
{\partial \chi _{1,2}}=0,                                         \label{III:7}
\end{equation}
where the effective potential
\begin{equation}
U=\frac{1}{2}\Omega^{2}\left(f_{-1}^{2}+f_{+1}^{2}\right)-V.      \label{III:8}
\end{equation}
The regularity of the soliton  field  configuration  and  the finiteness of the
soliton energy lead to the following boundary conditions:
\begin{subequations}                                              \label{III:9}
\begin{align}
f_{\pm 1}^{\prime }\left( 0\right) &=0, & f_{\pm 1}\left( r\right) & \underset
{r\rightarrow \infty }{\longrightarrow }0,                       \label{III:9a}
 \\
\chi _{1,2}^{\prime }\left( 0\right) &=0, & \chi _{1,2}\left( r\right)
& \underset{r\rightarrow \infty }{\longrightarrow }\chi_{1,2\,\text{vac}},
                                                                 \label{III:9b}
 \\
\Omega ^{\prime }\left( 0\right) &=0, & \Omega \left( r\right) & \underset{
r\rightarrow \infty}{\longrightarrow} \omega.                    \label{III:9c}
\end{align}
\end{subequations}

The boundary conditions  in  Eqs.~(\ref{III:9b})  and  (\ref{III:9c}) need some
explanation.
From Eqs.~(\ref{II:1}) and (\ref{II:4}),  it  follows that the classical vacuum
of model (\ref{II:1}) is
\begin{equation}
F_{\mu \nu} = 0,\quad
\phi_{\pm 1} = 0,\quad
\phi_{0} = \phi_{0\,\text{vac}},                                 \label{III:10}
\end{equation}
where $\phi _{0\,\text{vac}}$ is an arbitrary complex constant.
From Eq.~(\ref{III:10}), it  follows  that  model  (\ref{II:1}) has an infinite
number of vacua at the classical level,  as reflected in the boundary condition
in Eq.~(\ref{III:9b}).
All of these vacua are invariant  under  both the $U(1)$ gauge and $\mathcal{N}
= 1$ supersymmetry transformations.
According   to     the    non-renormalisation    theorems   \cite{gsr_npb_1979,
 seiberg_pl_1993}, this will also be true when perturbative quantum corrections
are taken into account.

Eqs.~(\ref{III:3}), (\ref{III:7}), and (\ref{III:8})  tell  us  that $\chi_{1}$
and $\chi_{2}$  satisfy  the  same  linear  homogeneous  differential equation,
while Eq.~(\ref{III:9b}) tells us  that  $\chi_{1}$  and $\chi_{2}$ satisfy the
same homogeneous boundary condition at $r = 0$.
It follows that the ratio $\chi_{2}(r)/\chi_{1}(r)$ does not depend on $r$, and
is equal to $\chi_{2\,\text{vac}}/\chi_{1\,\text{vac}}$.
The phase of the ansatz function $\varphi_{0}(r)=2^{-1/2}(\chi_{1}(r)+i\chi_{2}
(r))$ is therefore a constant.
However, from  Eqs.~(\ref{III:2})  and  (\ref{III:3}),  it follows that in this
case, the energy density and  the  Lagrangian  density  do  not  depend  on the
phase of $\varphi_{0}\left(r\right)$.
Without loss  of  generality,  we  can set this phase (and hence $\chi_{2}(r)$)
equal to zero.

The field  configurations  of  model  (\ref{II:1})  are  determined up to gauge
transformations.
In particular, the choice  of  ansatz (\ref{III:1}) is equivalent to the choice
of the radial gauge.
However, this gauge  does  not  fix the soliton field configuration completely;
to do this, we need to impose an additional condition $\Phi(\infty) = 0$, which
is equivalent to Eq.~(\ref{III:9c}).

The basic property of any non-topological  soliton is that it is an extremum of
the energy functional $E$  at  a fixed value of some Noether charge $Q_{N}$ (in
our case $E=4\pi \int\nolimits_{0}^{\infty }\mathcal{E}(r)r^{2}dr$ and $Q_{N}=4
\pi e^{-1}\int\nolimits_{0}^{\infty }j^{0}(r)r^{2}dr$).
This property results in the differential relation
\begin{equation}
dE/dQ_{N} = \Omega_{\infty},                                     \label{III:11}
\end{equation}
where $\Omega_{\infty} \equiv \Omega(\infty) = \omega - e \Phi(\infty)=\omega$.
Note that a similar relation  also holds for the electrically charged magnetic
monopoles \cite{loginov_plb_822}.

Eqs.~(\ref{III:3}) and (\ref{III:8}) tell  us  that  the potentials $V$ and $U$
are invariant under the permutation $f_{-1}\leftrightarrow f_{+1}$.
It follows that if $f_{-1}(r)$, $f_{+1}(r)$, $\chi_{1}(r)$,  and $\Omega(r)$ is
a  solution  of   system  (\ref{III:5})  --  (\ref{III:7}),  then  $f_{+1}(r)$,
$f_{-1}(r)$, $\chi_{1}(r)$, and $\Omega(r)$ is also a solution.
Using qualitative research methods  for differential equations, it can be shown
that the solutions $f_{-1}(r)$ and $f_{+1}(r)$ coincide when the gauge coupling
constant $e = 0$.
In the following, we define  the function  $\delta\left(r,e^{2}\right) = f_{+1}
\left(r,e^{2}\right) - f_{-1}\left(r,e^{2}\right)$, where the dependence on the
gauge coupling constant is explicitly indicated and  we  use the  fact that the
potential $V$ depends on $e$ only through $e^{2}$.
The function $\delta\left(r,e^{2}\right)$  satisfies the nonlinear differential
equation
\begin{flalign}
&\delta ^{\prime \prime }+\frac{2}{r}\delta ^{\prime }+\left[ \Omega
^{2}+2^{-1}g^{2}\left(f_{-1}^{2}-\chi_{1}^{2}\right)-2e^{2}f_{-1}^{2}\right]
\delta                                                              \nonumber
  \\
&+2^{-1}\left(g^{2}-4e^{2}\right)f_{-1}\delta^{2}
-2^{-1}e^{2}\delta^{3} = 0,                                      \label{III:12}
\end{flalign}
where the dependence of $\delta$, $\Omega$, $f_{-1}$, and $\chi_{1}$ on $r$ and
$e^{2}$ is omitted.
From Eq.~(\ref{III:9a}), it follows that $\delta\left(r,e^{2}\right)$ satisfies
the boundary conditions
\begin{equation}
\delta ^{\prime }\left( 0,e\right) = 0, \quad \delta \left(
\infty ,e\right) = 0.                                            \label{III:14}
\end{equation}

Our goal is   to   find   the  derivatives  $\delta^{(n)}  \equiv  \partial^{n}
\delta/\partial e^{n}$ at $e = 0$.
To do this, we differentiate  Eq.~(\ref{III:12})  with respect to $e$, and then
set $e = 0$.
As     a     result,     we     obtain     the     trivial    linear   equation
$\delta^{\left(1\right)\prime\prime}+2r^{-1}\delta^{\left(1\right)\prime} = 0$.
Its solution  must  satisfy   the  boundary   conditions  in Eq.~(\ref{III:14})
differentiated with respect to $e$, and it  is  therefore easy  to see that the
solution is $\delta^{\left(1\right)}(r,0) = 0$.
Thus, we have established that $\delta(r, 0) = 0$  and $\delta^{\left(1\right)}
(r,0) = 0$.
By continuing to differentiate  Eq.~(\ref{III:12}) with respect to $e$, setting
$e = 0$, and taking into account  the  previous  results  at  each step, it can
be shown that $\delta^{(n)}(r, 0) = 0$ for any $n \ge 0$.
It follows that $\delta\left(r, e^{2}\right)$ vanishes, and hence $f_{+1}\left(
r,e^{2}\right) = f_{-1}\left(r,e^{2}\right) \equiv f\left(r,e^{2}\right)$.

We now  examine  the  asymptotics   of   the   soliton  fields  for  large $r$.
Suppose that  $f(r)$  tends  to  zero  exponentially as $r \rightarrow \infty$.
In this case,  we  can  neglect  the  nonlinear terms in Eqs.~(\ref{III:5}) and
(\ref{III:7}), and obtain the asymptotic forms of $\Omega(r)$ and $\chi_{1}(r)$
as $r \rightarrow \infty$:
\begin{equation}
\Omega \sim \omega -\frac{e}{4\pi}\frac{Q}{r},                   \label{III:15}
\end{equation}
\begin{equation}
\chi _{1}\sim \chi _{1\,\text{vac}}-\frac{1}{4\pi }\frac{Q_{\text{s}}}{r},
                                                                 \label{III:16}
\end{equation}
where $Q = 4 \pi \int\nolimits_{0}^{\infty }j^{0}(r)r^{2}dr$  is  the  electric
charge of the soliton, and $Q_{s}$ is the scalar charge defined by analogy with
the large-distance asymptotics $\Phi\sim Q/(4\pi r)$ for the electric potential.
We see that both $\Omega=\omega-e\Phi$ and $\chi_1$ tend rather slowly ($\propto
r^{-1}$) to their limiting values as $r \rightarrow \infty$.
It should be noted that nontopological solitons with  a long-range scalar field
were studied in Refs.~\cite{lgn_pan_2010, lr_mpla_2011, lpshnir_2018}.
Furthermore, electrically  charged  nontopological  solitons  with a long-range
scalar   field   were    studied    in    Refs.~\cite{lshnir_2019, lshnir_2022,
 kunzlsh_2022}.

By substituting Eqs.~(\ref{III:15})  and  (\ref{III:16})  into  Eq.~(\ref{III:6}),
retaining the terms linear  in $f(r)$,  and  solving the resulting differential
equation, we obtain the large-distance asymptotics of $f(r)$ as
\begin{flalign}
f\left( r\right) & \sim f_{\infty}e^{-\Delta r}\left(\Delta r\right)^{\beta}
                                                                    \nonumber
 \\
& \times \left(1-\frac{a^{2}}{32\pi^{2}\Delta^{3}r}
-\frac{b}{8\pi\Delta^{2}r}\right),                               \label{III:17}
\end{flalign}
where $f_{\infty}$ is a constant,
\begin{subequations}                                             \label{III:18}
\begin{flalign}
\Delta  &= \left( \omega _{\max }^{2}-\omega^{2}\right)^{1/2},  \label{III:18a}
 \\
a &= e\omega_{\max}\left\vert Q\right\vert
-g\left\vert\omega\right\vert\left\vert Q_{\text{s}}\right\vert,\label{III:18b}
 \\
b &= e \left\vert \omega \right\vert \left\vert Q \right\vert
- g \omega_{\max} \left\vert Q_{\text{s}} \right\vert,          \label{III:18c}
 \\
\beta  &= -1-b/\left( 4\pi \Delta \right),                      \label{III:18d}
\end{flalign}
\end{subequations}
and the parameter  $\omega_{\max} = 2^{-1/2} g \left\vert \chi_{1\, \text{vac}}
\right\vert$.
We see that our assumption about the exponential asymptotics  of  $f(r)$ turned
out to be correct; we also see that the long-range terms in  the asymptotics of
$\Omega(r)$ and  $\chi_{1}(r)$   modify   the   pre-exponential  factor  in the
asymptotics of $f(r)$.
Furthermore, we can conclude that the nontopological soliton cannot  exist when
$\left\vert \omega \right\vert > \omega_{\max}$, since in this case asymptotics
(\ref{III:17})  shows  oscillating  behavior, leading to an infinite energy and
charge  for the corresponding field configuration.

The presence of two long-range fields in Eqs.~(\ref{III:15}) and (\ref{III:16})
leads to a modification  of  the  intersoliton  interaction  potential at large
distances.
It can be shown that in the case  of  large  distances  and low velocities, the
leading term of the intersoliton interaction potential is
\begin{equation}
V_{12}=\frac{Q^{\left( 1\right) }Q^{\left( 2\right) }-Q_{\text{s}}^{\left(
1\right) }Q_{\text{s}}^{\left( 2\right) }}{4\pi r_{12}},         \label{III:19}
\end{equation}
where $Q^{\left( i \right)}$ ($Q_{\text{s}}^{\left( i\right)}$) is the electric
(scalar) charge of the $i$-th soliton.
Eq.~(\ref{III:19}) tells us that the energy  of the intersoliton interaction is
the sum of the energies of the Coulomb and scalar interactions.
Depending on  the  signs  of $Q^{(1)}$ and $Q^{(2)}$, the Coulomb energy may be
both positive (repulsion) and negative (attraction).
At the same time, it follows from the inhomogeneity  of the  boundary condition
in Eq.~(\ref{III:9b}) that  for  the  fixed  vacuum  in Eq.~(\ref{III:10}), the
scalar charges $Q_{\text{s}}^{\left( i \right) }$  of  the  solitons  must have
the same sign.
Hence, unlike the  Coulomb  field,  the long-range scalar field always leads to
attraction between solitons.

\section{Fermionic zero modes}
\label{seq:IV}

The Lagrangian  density  (\ref{II:1})  is  written  in  the  Wess-Zumino gauge,
meaning  that  the  corresponding  action  $S  =  \int\mathcal{L}d^{4}x$ is not
invariant under the usual $\mathcal{N} = 1$ supersymmetry transformations.
However, it will be invariant under  the modified supersymmetry transformations
\cite{de_Witt_Freedman_1975}:
\begin{subequations}                                               \label{IV:1}
\begin{flalign}
\delta \phi _{n} &= \sqrt{2}\overline{\alpha _{R}}\psi _{nL},     \label{IV:1a}
 \\
\delta \psi _{nL} &= \sqrt{2}\gamma ^{\mu }\left( D_{\mu }\phi _{n}\right)
\alpha _{R}+\sqrt{2}\mathcal{F}_{n}\alpha _{L},                   \label{IV:1b}
 \\
\delta A_{\mu } &= \bar{\alpha }\gamma _{\mu }\lambda,            \label{IV:1c}
 \\
\delta \lambda  &= i \mathcal{D}\gamma_{5}\alpha - \frac{1}{4} F_{\mu \nu}
\left[\gamma^{\mu},\gamma^{\nu}\right] \alpha,                    \label{IV:1d}
\end{flalign}
\end{subequations}
where
\begin{equation}
\alpha =-i%
\begin{pmatrix}
\epsilon _{a} \\
\sum\nolimits_{b}e_{ab}\epsilon_{b}^{\ast}
\end{pmatrix},                                                     \label{IV:2}
\end{equation}
\begin{equation}
%\epsilon =%
\begin{pmatrix}
\epsilon _{1} \\
\epsilon _{2}%
\end{pmatrix}%
=%
\begin{pmatrix}
\epsilon _{11}+i\epsilon _{12} \\
\epsilon _{21}+i\epsilon _{22}%
\end{pmatrix},                                                     \label{IV:3}
\end{equation}
\begin{equation}
\mathcal{F}_{n}=-\left(\partial f/\partial\phi_{n}\right)^{\ast},  \label{IV:4}
\end{equation}
and
\begin{equation}
\mathcal{D}=e\left(\phi_{+1}^{\ast}\phi_{+1}-\phi_{-1}^{\ast}\phi_{-1}\right).
                                                                   \label{IV:5}
\end{equation}
In Eq.~(\ref{IV:3}),  $\epsilon_{i j}$  are  real  infinitesimal  anticommuting
transformation parameters and $e_{a b}$  is an antisymmetric $2\times 2$ matrix
with $e_{1 2}= +1$, from which it follows that $\alpha$ in  Eq.~(\ref{IV:2}) is
the Majorana spinor.
In Eq.~(\ref{IV:4}), the  auxiliary  fields  $\mathcal{F}_{n}$ are expressed in
terms of superpotential  (\ref{II:3}), and  it  is  assumed that all the fields
in Eqs.~(\ref{IV:1a})--(\ref{IV:1d})       satisfy         field      equations
(\ref{II:5})--(\ref{II:8}).

Fermionic  zero  modes  are   generated   by   the  action  of  transformations
(\ref{IV:1b}) and  (\ref{IV:1d})   on   purely   bosonic   field  configuration
(\ref{III:1}).
To represent these in a compact form,  we  introduce a column $\Psi$ consisting
of four fermionic fields included in the Lagrangian (\ref{II:1}).
The transposed form of $\Psi$ is
\begin{equation}
\Psi^{\text{T}} = N\left(\psi_{+1 L}^{\text{T}},\psi_{0 L}^{\text{T}},
\psi_{-1 L}^{\text{T}},\lambda^{\text{T}}\right),                  \label{IV:6}
\end{equation}
where
\begin{equation}
\psi _{\pm 1 L}=
\begin{pmatrix}
A_{\pm 1}f+Cf^{\prime } \\
B_{\pm 1}f+Df^{\prime } \\
0 \\
0%
\end{pmatrix}
e^{\mp i\omega t},                                                 \label{IV:7}
\end{equation}
\begin{equation}
\psi _{0 L}=%
\begin{pmatrix}
i\epsilon _{1}2^{-\frac{1}{2}}gf^{2}+C\chi _{1}^{\prime } \\
i\epsilon _{2}2^{-\frac{1}{2}}gf^{2}+D\chi _{1}^{\prime } \\
0 \\
0%
\end{pmatrix},                                                     \label{IV:8}
\end{equation}
\begin{equation}
\lambda =i\Phi ^{\prime }%
\begin{pmatrix}
\epsilon _{1}c+\epsilon _{2}e^{-i\varphi }s \\
-\epsilon _{2}c+\epsilon _{1}e^{i\varphi }s \\
-\epsilon _{2}^{\ast }c+\epsilon _{1}^{\ast }e^{-i\varphi }s \\
-\epsilon _{1}^{\ast }c-\epsilon _{2}^{\ast }e^{i\varphi }s%
\end{pmatrix},                                                     \label{IV:9}
\end{equation}
and $N$ is a normalisation factor.
For brevity, in Eqs.~(\ref{IV:7})--(\ref{IV:9}), we use the notation
\begin{subequations}                                              \label{IV:10}
\begin{flalign}
A_{\pm 1} & =\pm i\epsilon _{2}^{\ast }\Omega +i2^{-\frac{1}{2}}\epsilon
_{1}g\chi _{1}, \\
B_{\pm 1} & =\mp i\epsilon _{1}^{\ast }\Omega +i2^{-\frac{1}{2}}\epsilon
_{2}g\chi _{1}, \\
C &= -\epsilon _{2}^{\ast }c+\epsilon _{1}^{\ast }e^{-i\varphi }s, \\
D &= -\epsilon _{1}^{\ast }c-\epsilon _{2}^{\ast }e^{i\varphi }s,
\end{flalign}
\end{subequations}
where $c = \cos(\theta)$,  $s = \sin(\theta)$,  $\epsilon_{1} = \epsilon_{11} +
i\epsilon_{12}$,  and  $\epsilon_{2} = \epsilon_{21} + i \epsilon_{22}$.

Eqs.~(\ref{IV:7})--(\ref{IV:9})  depend  linearly  on  the  four  anticommuting
parameters $\epsilon_{ij}$, and hence  Eq.~(\ref{IV:6}) can be written as $\Psi
= \sum\nolimits_{ij} \epsilon_{ij} \Psi_{ij}$.
It follows  that  there  are  four (according to the number of the $\mathcal{N}
= 1$ supersymmetry generators)  independent  fermionic  zero  modes $\Psi_{ij}$
expressed in terms of ansatz functions (\ref{III:1}).
It can be shown that the  components  of  the  fermionic zero modes $\Psi_{ij}$
satisfy  field    equations   (\ref{II:7})   and   (\ref{II:8}),  provided that
the   ansatz     functions    $\Omega$,    $f$,    and    $\chi_{1}$    satisfy
Eqs.~(\ref{III:5})--(\ref{III:7}).
The fermionic zero modes satisfy the orthonormality condition
\begin{equation}
\int \Psi_{ij}^{\dagger}\Psi_{i^{\prime}j^{\prime }}d^{3}x =
\delta_{i i^{\prime}}\delta_{j j^{\prime}},                       \label{IV:11}
\end{equation}
provided that the normalisation factor
\begin{flalign}
N &= \left[ 2\pi \int\nolimits_{0}^{\infty }\left[4\left(\Phi^{\prime
2}+f^{\prime 2}\right) +2\chi _{1}^{\prime 2}\right. \right.        \nonumber
  \\
&\biggl. \left. +g^{2}f^{4}+2f^{2}\left( 2\Omega ^{2}+g^{2}\chi
_{1}^{2}\right) \right] r^{2}dr\biggr]^{-\frac{1}{2}}.            \label{IV:12}
\end{flalign}

From Eq.~(\ref{IV:9}), it follows that  the  gaugino component $\lambda$ of the
fermionic zero mode $\Psi_{ij}$ is proportional to the  electric field strength
$E_{r} =-\Phi'$ of the soliton, and therefore decreases rather slowly ($\propto
r^{-2}$) at large distances.
Furthermore,  Eqs.~(\ref{III:16})  and   (\ref{IV:8})  tell  us  that  at large
distances, the  component $\psi_{0 L}\propto \chi_{1}' \sim Q_{\text{s}}/(4 \pi
r^2)$.
We see that similarly  to  the  $\lambda$ component, the $\psi_{0 L}$ component
of $\Psi_{i j}$ decreases slowly ($\propto r^{-2}$) at large distances.
In contrast,  Eqs.~(\ref{III:17})   and   (\ref{IV:7})  tell  us  that  the two
remaining components $\psi_{\pm 1 L}$  of $\Psi_{i j}$, which correspond to the
short-range scalar fields $\phi_{\pm 1}$, decrease  exponentially away from the
soliton.

Written in terms of  the  left-handed  fermion  fields  (including the massless
``neutrino'' $\psi_{0 L}$),  the Lagrangian (\ref{II:1}) is not invariant under
the $P$ and $C$ transformations; it is, however,  invariant  under the combined
$CP$ transformation.
Under the latter transformation, the original soliton solution ($f(r)\exp(\mp i
\omega t)$,  $\chi_{1}(r)$,  $\Phi(r)$,  $\Omega(r)$)  of   the  energy $E$ and
electric charge $Q$ is transformed  into an antisoliton solution ($f(r)\exp(\pm
i \omega t)$,  $\chi_{1}(r)$,  $-\Phi(r)$, $-\Omega(r)$)  of the energy $E$ and
electric charge $-Q$.
It can be shown that under  the  $CP$  transformation, the fermionic zero modes
$\Psi_{ij}$  of  the  soliton   turn  into  those  $\tilde{\Psi}_{ij}$  of  the
antisoliton:
\begin{eqnarray}
\left[\Psi_{11}(x)\right]^{CP} &=&-\tilde{\Psi}_{22}(x),            \nonumber
 \\
\left[\Psi_{12}(x)\right]^{CP} &=&-\tilde{\Psi}_{21}(x),            \nonumber
 \\
\left[\Psi_{21}(x)\right]^{CP} &=&\tilde{\Psi}_{12}(x),             \nonumber
 \\
\left[\Psi_{22}(x)\right]^{CP} &=&\tilde{\Psi}_{11}(x).           \label{IV:14}
\end{eqnarray}
This is  because  the  $CP$   transformation  is  a  discrete  symmetry  of the
Lagrangian (\ref{II:1}), and  hence  must  convert one fermion-soliton solution
into another.

\section{Numerical results}
\label{seq:V}

The  system  of   differential   equations  (\ref{III:5}) -- (\ref{III:7}) with
boundary conditions (\ref{III:9}) represents a  mixed  boundary  value  problem
on the semi-infinite interval $r\in\left[0,\infty\right)$.
To solve this system, we use the numerical methods provided in the {\sc{Maple}}
package \cite{maple}.

Formally, the boundary value  problem (\ref{III:5}) -- (\ref{III:9}) depends on
five parameters: $\omega$, $m$, $g$, $e$, and $\chi_{1\,\text{vac}}$.
However, it is easily shown that the  energy  and Noether charge of the soliton
depends nontrivially on only three dimensionless parameters:
\begin{flalign}
& E\left( \omega ,m,g,e,\chi_{1\,\text{vac}}\right) = mg^{-2}\tilde{E}
\left(\tilde{\omega},\tilde{e},\tilde{\chi}_{1\,\text{vac}}\right), \label{V:1}
 \\
& Q_{N}\left( \omega ,m,g,e,\chi_{1\,\text{vac}}\right)  = g^{-2}\tilde{Q}
_{N}\left(\tilde{\omega},\tilde{e},\tilde{\chi}_{1\,\text{vac}}
\right),                                                            \label{V:2}
\end{flalign}
where $\tilde{\omega} = \omega/m$, $\tilde{e}  =  e/g$,  and $\tilde{\chi}_{1\,
\text{vac}}=\chi_{1\,\text{vac}}/m$.
Hence, without loss  of  generality,  we  can  set  the  parameters $m$ and $g$
equal to unity.
In addition, we set the dimensionless parameter $\tilde{\chi}_{1\,\text{vac}} =
2\sqrt{2}$ in these numerical calculations.

\begin{figure}[tbp]
\includegraphics[width=7.8cm]{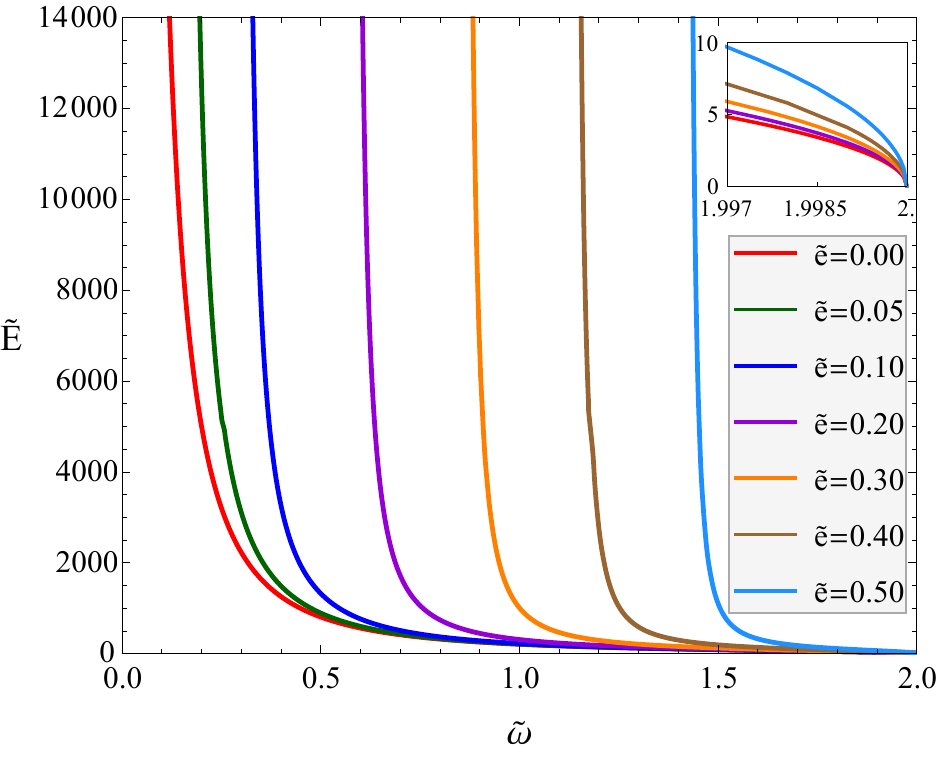}
\caption{\label{fig1}   Dependence  of  the  soliton  energy $\tilde{E}$ on the
phase frequency $\tilde{\omega}$  for  several  values  of  the  gauge coupling
constant $\tilde{e}$.}
\end{figure}

Figure~\ref{fig1} shows the dependence of the soliton energy $\tilde{E}$ on the
phase frequency  $\tilde{\omega}$  for  several  values  of  the gauge coupling
constant $\tilde{e}$.
We see that for each $\tilde{e}$, the phase frequency $\tilde{\omega}\in \left(
\tilde{\omega}_{\min}(\tilde{e}), \tilde{\omega}_{\max}\right]$, where $\tilde{
\omega}_{\max} = 2^{-1/2}\tilde{\chi}_{1\, \text{vac}} = 2$.
As $\tilde{e}$ decreases, the minimum allowable frequency $\tilde{\omega}_{\min
}(\tilde{e})$ falls  monotonically,  reaching the limiting value $\tilde{\omega
}_{\min}(0) = 0$.
Using numerical  methods,  we  can  show  that  as  $\tilde{\omega} \rightarrow
\tilde{\omega}_{\min}(\tilde{e})$, the soliton energy
\begin{equation}
\tilde{E}\left(\tilde{\omega},\tilde{e}\right) \sim a(\tilde{e})
(\tilde{\omega}-\tilde{\omega}_{\min}(\tilde{e}))^{-2},             \label{V:3}
\end{equation}
where $a(\tilde{e})$ is a function of $\tilde{e}$.
It follows that the  soliton  energy  increases indefinitely as $\tilde{\omega}
\rightarrow \tilde{\omega}_{\min}(\tilde{e})$.
On the  other  hand, $\tilde{\omega}_{\min}(\tilde{e})$ monotonically increases
with $\tilde{e}$, meaning that  there  is  a  limiting value $\tilde{e}_{\max}$
for which  $\tilde{\omega}_{\min}(\tilde{e}_{\max})  =  \tilde{\omega}_{\max}$.
It follows that the nontopological  soliton  can exist only when $\tilde{e} \in
\left[0, \tilde{e}_{\max}\right)$.

In the subplot in Fig.~\ref{fig1},  we  can  see  the curves $\tilde{E}(\tilde{
\omega},\tilde{e})$ in the vicinity of the maximum   allowable  phase frequency
$\tilde{\omega}_{\max}$. %$ = 2^{-1/2} \tilde{\chi}_{1\, \text{vac}} = 2$.
All  the   curves   $\tilde{E}(\tilde{\omega},  \tilde{e})$ in the subplot tend
to zero as $\tilde{\omega} \rightarrow \tilde{\omega}_{\max}$.
It has been found numerically that as $\tilde{\omega} \rightarrow \tilde{\omega
}_{\min}(\tilde{e})$, the soliton energy
\begin{equation}
\tilde{E}\left(\tilde{\omega},\tilde{e}\right)\approx b(\tilde{e}
)\left(\tilde{\omega}_{\max}-\tilde{\omega}\right)^{1/2},           \label{V:4}
\end{equation}
where $b(\tilde{e})$ is an increasing function of $\tilde{e}$.

According  to  Eq.~(\ref{III:11}),  the  curves  $\tilde{Q}_{N}(\tilde{\omega},
\tilde{e})$ are related to the curves $\tilde{E}(\tilde{\omega}, \tilde{e})$ by
the  integral  relation  $\tilde{Q}_{N}\left( \tilde{\omega }, \tilde{e}\right)
=-\int\nolimits_{\tilde{\omega}}^{\tilde{\omega}_{\max}}\tau^{-1}\partial_{\tau
}\tilde{E}\left( \tau ,\tilde{e}\right) d\tau$.
It follows  that   the  curves  $\tilde{Q}_{N}(\tilde{\omega}, \tilde{e})$ will
be similar to  the  curves   $\tilde{E}(\tilde{\omega},  \tilde{e})$  shown  in
Fig.~\ref{fig1}; in  particular, the  behavior  of  the  curves $\tilde{Q}_{N}(
\tilde{\omega},\tilde{e})$ in the neighborhoods of $\tilde{\omega}_{\min}$  and
$\tilde{\omega}_{\max}$ is the same as  that  of  the curves $\tilde{E}(\tilde{
\omega}, \tilde{e})$.

\begin{figure}[tbp]
\includegraphics[width=7.8cm]{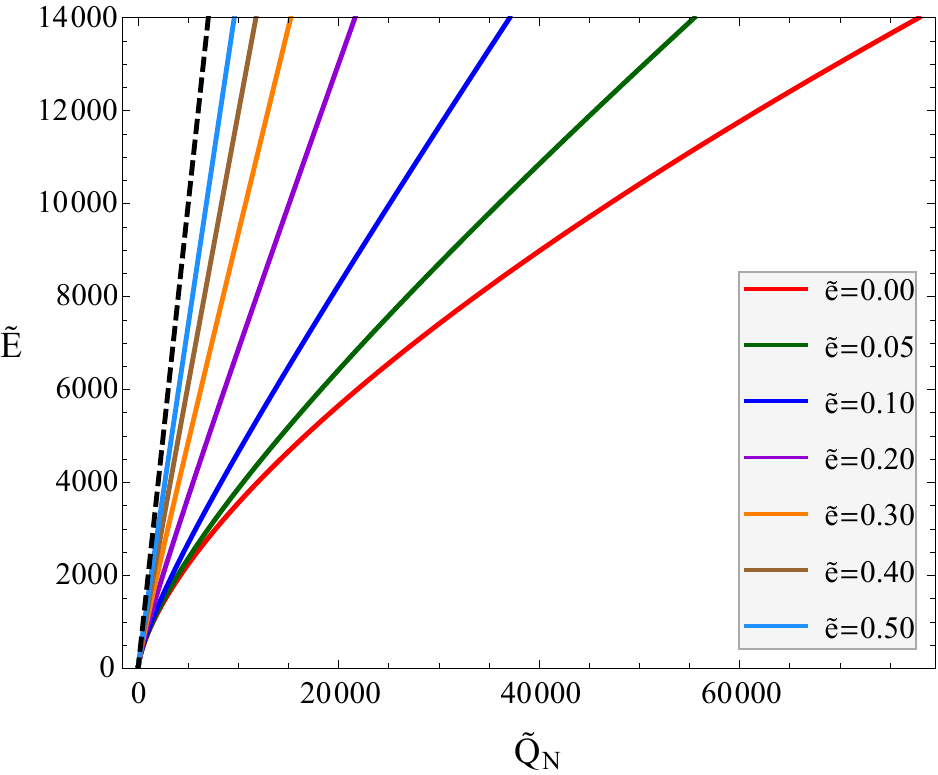}
\caption{\label{fig2}     Dependence  of  the  soliton  energy  $\tilde{E}$  on
the Noether  charge  $\tilde{Q}_{N}$  for  several values of the gauge coupling
constant $\tilde{e}$.}
\end{figure}

Figure~\ref{fig2} shows the  dependence  of  the  soliton energy $\tilde{E}$ on
the Noether  charge  $\tilde{Q}_{N}$  for  several values of the gauge coupling
constant $\tilde{e}$.
In Fig.~\ref{fig2}, the  black  dashed  line $\tilde{E} = \tilde{\omega}_{\max}
\tilde{Q}_{N}$ corresponds to the energy  of  a  plane-wave  configuration with
a given Noether charge $\tilde{Q}_{N}$.
We see that for all values of $\tilde{e}$ considered here,  the energies of the
solitons with  a  given  $\tilde{Q}_{N}$  are  lower  than  the  energy  of the
corresponding plane-wave configuration.
It follows that these solitons are stable  against  decay  into massive charged
$\phi$-mesons.
%corresponding to the complex scalar fields $\phi_{\pm}$. %scalar bosons.

\begin{figure}[tbp]
\includegraphics[width=7.8cm]{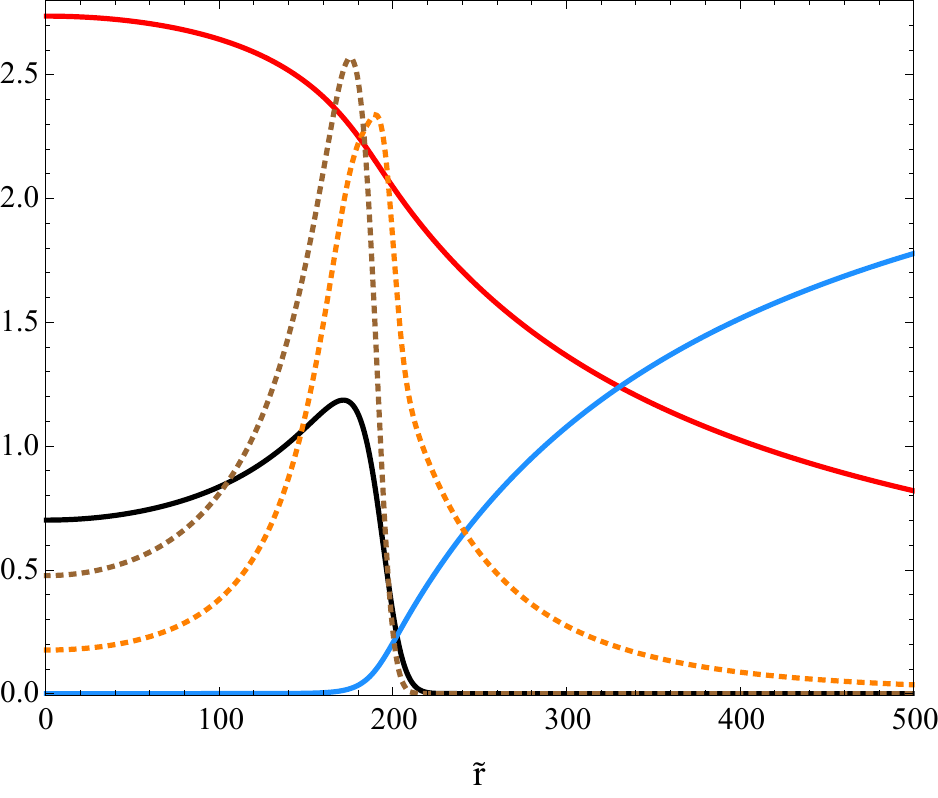}
\caption{\label{fig3}    Scaled  dimensionless  functions  $10 \times \tilde{f}
(\tilde{r})$ (solid black), $\tilde{\Phi}(\tilde{r})$ (solid red), $\tilde{\chi
}_{1}(\tilde{r})$  (solid blue), $10^{4} \times \tilde{\mathcal{E}}(\tilde{r})$
(dashed orange), and $10^{3}\times\tilde{j}_{N}^{0}(\tilde{r})$ (dashed brown).
The functions correspond to the parameters $\tilde{e}=0.1$ and  $\tilde{\omega}
= 0.32214$.}
\end{figure}

We have established that  the  energy $\tilde{E}(\tilde{\omega},\tilde{e})$ and
the Noether  charge  $\tilde{Q}_{N}(\tilde{\omega}, \tilde{e})$  of the soliton
increase  indefinitely  as  $\tilde{\omega}  \rightarrow  \tilde{\omega}_{\min}
(\tilde{e})$.
In view of this, it would be interesting to explore the behavior of the soliton
fields in this limit.
To do this, we define the dimensionless profile functions $\tilde{f}(\tilde{r})
=  m^{-1} g f(r)$, $\tilde{\chi}_{1}(\tilde{r})  =  m^{-1}  g \chi_{1}(r)$, and
$\tilde{\Phi}(\tilde{r})  =  m^{-1}  g  \Phi(r)$, where $\tilde{r} = m^{-1} r$.
We also define the dimensionless energy density $\tilde{\mathcal{E}}(\tilde{r})
= m^{-4} g^{2} \mathcal{E}(r)$  and  the  dimensionless  Noether charge density
$\tilde{j}_{N}^{0}(\tilde{r}) = m^{-3} g^{2} j_{N}^{0}(r)$.
Figure~\ref{fig3} shows  these  dimensionless  functions  for  parameter values
$\tilde{e} = 0.1$ and $\tilde{\omega} = 0.32214$.
Note that $\tilde{\omega}=0.32214$ is the minimum value of the phase frequency,
which we were able  to  achieve  by  numerical  methods  for $\tilde{e} = 0.1$.
We see that only $\tilde{f}(\tilde{r})$  and $\tilde{j}^{0}_{N}(\tilde{r})$ are
localised,  whereas  $\tilde{\Phi}(\tilde{r})$,  $\tilde{\chi}_{1}(\tilde{r})$,
and $\tilde{\mathcal{E}}(\tilde{r})$   are  long-range,   which  is  consistent
with the  asymptotic   forms   in    Eqs.~(\ref{III:15}),   (\ref{III:16}), and
(\ref{III:17}).
We also see that $\tilde{\chi}_{1}(\tilde{r}) \approx 0$ in the interior of the
soliton.
The long-range character ($\propto r^{-4}$)   of   the  energy density $\tilde{
\mathcal{E}}$ arises from  the  gradient  of  the long-range electric potential
$\tilde{\Phi}$  and  the  gradient  of  the  long-range  neutral  scalar  field
$\tilde{\chi}_{1}$.
According to Eq.~(\ref{III:4}),  the  local  character  of  the  charge density
$\tilde{j}_{N}^{0}$ is due to the local character  of the function $\tilde{f}$.
Note that the electrostatic  repulsion  causes  the  electric charge density to
increase near the surface of the soliton.

Eq.~(\ref{III:16}) tells  us  that  the  asymptotics  of  $\tilde{\chi}_{1}$ is
characterised  by  the  scalar  charge $\tilde{Q}_{\text{s}} = g Q_{\text{s}}$.
Using numerical methods, we find  that  similarly to the energy $\tilde{E}$ and
the Noether charge $\tilde{Q}_{N}$, the scalar charge
\begin{equation}
\tilde{Q}_{\text{s}}\left(\tilde{\omega},\tilde{e}\right)
\propto \left(\tilde{\omega}-\tilde{\omega}_{\min}
\left(\tilde{e}\right)\right)^{-2}                                  \label{V:5}
\end{equation}
as $\tilde{\omega} \rightarrow \tilde{\omega}_{\min}$.
However,  unlike  the  Noether  (electric)  charge $Q_{N}$ ($Q = e Q_{N}$), the
scalar charge $Q_{\text{s}}$ is simply  a definition  and is not related to any
symmetry of model (\ref{II:1}).

\section{Conclusion}
\label{seq:VI}

In the present paper,  we  show  that  an  electrically  charged nontopological
soliton exists in a  version  of  $\mathcal{N}  =  1$   supersymmetric   scalar
electrodynamics.
A characteristic feature of  this  soliton is  the presence of  two  long-range
fields, which slowly ($ \propto r^{-1}$) tend to limiting values: these are the
electrostatic Coulomb  field,  and  the  electrically  neutral  massless scalar
field.
The presence of these  two  long-range  fields  leads  to a modification of the
intersoliton interaction in comparison with the purely Coulomb case.
Another feature of the soliton  is  that  its  energy  and electric charge take
arbitrarily large  values when  the modulus of the phase frequency tends to the
minimum possible value.
In contrast, the energy and electric  charge  of  the  soliton  vanish when the
modulus of the phase frequency tends to the maximum possible value.

We note that in   the   general   case,   the   energy and electric charge of a
nontopological soliton cannot  be  arbitrarily  large  due to Coulomb repulsion
\cite{klee, gulamov_2015}.
We avoid this restriction  because  the  attraction  due to the massless scalar
field compensates for the Coulomb repulsion.
A similar  situation  also  arises   in   the   massless  limit  of  the gauged
Fridberg-Lee-Sirlin model \cite{lshnir_2019, lshnir_2022}.
It is also worth  noting  that  the  electric  charge  and  energy  of the dyon
(electrically  charged  magnetic  monopole) also cannot be arbitrarily large in
the general non-BPS case \cite{bkt_1999}.
Only in the BPS limit, when the scalar  field of the dyon becomes massless, can
the energy and electric charge take arbitrarily large values.

The $\mathcal{N}=1$ supersymmetry of  the  model  makes  it  possible to obtain
expressions for the  fermionic  zero  modes  in  terms of bosonic fields of the
soliton.
The fermionic  zero  modes  are  bound  states  of  the fermion-soliton system,
and their components that correspond  to the long-range bosonic fields are also
long-range.
In accordance with the  number of $\mathcal{N}=1$ supersymmetry generators, the
number of independent fermionic zero modes of the soliton is four.
The fermionic zero modes of two  solitons  with  opposite  electric charges are
related by the $CP$ transformation.

In this work, we have investigated  a nontopological soliton of an $\mathcal{N}
= 1$ supersymmetric Abelian gauge model.
It is  known \cite{fried1, fried2},  however,  that nontopological solitons can
also exist in non-Abelian gauge models.
In particular, it was shown in  Ref.~\cite{lgn_jetp_2012}  that an electrically
charged nontopological   soliton   exists   in   the   Weinberg-Salam  model of
electroweak interactions.
This model  allows  for  $\mathcal{N} = 1$  supersymmetric  extension,  and its
fermionic sector  contains  both  massive ($e$,  $\mu$,  $\tau$)  and  massless
($\nu_{e}$, $\nu_{\mu}$, $\nu_{\tau}$) fermions.
The bosonic superpartners of the  neutrinos (sneutrinos) also have zero masses.
We   can   assume     that,    similarly    to    the    nonsupersymmetric case
\cite{lgn_jetp_2012},  an  electrically   charged  nontopological  soliton also
exists in this model, meaning  that  some  properties  of  this soliton will be
similar to those studied in this work.
In particular, in addition  to  the long-range Coulomb field, this soliton will 
have long-range fields of massless sneutrinos.
Furthermore, it will be  possible  to  express the fermionic zero modes of this
soliton in terms of its bosonic fields.

\section*{Acknowledgements}

This work was supported by the Russian Science Foundation, grant No 23-11-00002.




%\section*{References}

\bibliographystyle{elsarticle-num}

\bibliography{article}




%%\bibitem{}

%%\end{thebibliography}

\clearpage

\section*{Figure captions}

Fig.~1.     Dependence of the soliton energy $\tilde{E}$ on the phase frequency
$\tilde{\omega}$ for several values of the gauge coupling constant $\tilde{e}$.

\vspace{5mm}

Fig.~2.      Dependence  of  the  soliton  energy  $\tilde{E}$  on  the Noether
charge  $\tilde{Q}_{N}$  for  several  values  of  the  gauge coupling constant
$\tilde{e}$.

\vspace{5mm}

Fig.~3.      Scaled     dimensionless     functions     $10   \times  \tilde{f}
(\tilde{r})$ (solid black), $\tilde{\Phi}(\tilde{r})$ (solid red), $\tilde{\chi
}_{1}(\tilde{r})$  (solid blue), $10^{4} \times \tilde{\mathcal{E}}(\tilde{r})$
(dashed orange), and $10^{3}\times\tilde{j}_{N}^{0}(\tilde{r})$ (dashed brown).
The functions correspond to the parameters $\tilde{e}=0.1$ and  $\tilde{\omega}
= 0.32214$.

\end{document}